\documentclass[conference]{IEEEtran}
\usepackage{fancyhdr}
\usepackage{amsmath,graphicx}
\usepackage{color}
\usepackage{graphicx}
\usepackage{epstopdf}
\usepackage{amsmath}
\usepackage{amssymb}
\usepackage{mathrsfs}
\usepackage{balance}
\usepackage{tikz}
\usepackage{bm}
\usepackage[english]{babel}
\usepackage{cite}
\usepackage{rotfloat}
\usepackage{mathtools}
\usepackage[font=normalsize,labelfont=bf]{caption}
\usepackage{amsmath}
\usepackage{makecell}
\usepackage{algorithm,algorithmic}
\usepackage{multirow}
\usepackage{subfigure}
\usepackage{booktabs}
\usepackage{colortbl}
\usepackage{multirow}% http://ctan.org/pkg/multirow
\usepackage{hhline}% http://ctan.org/pkg/hhline
\usepackage{stfloats}% <-- added
\usepackage{multicol}
\usepackage{bbm}
\usepackage{cases}
\graphicspath{ {Figures/} }
\makeatletter
\newcommand\fs@betterruled{%
  \def\@fs@cfont{\bfseries}\let\@fs@capt\floatc@ruled
  \def\@fs@pre{\vspace*{5pt}\hrule height.8pt depth0pt \kern2pt}%
  \def\@fs@post{\kern2pt\hrule\relax}%
  \def\@fs@mid{\kern2pt\hrule\kern2pt}%
  \let\@fs@iftopcapt\iftrue}
\floatstyle{betterruled}
\restylefloat{algorithm}
\makeatother

\newcommand{\T}{{\scriptscriptstyle\mathsf{T}}}
\renewcommand{\H}{{\scriptscriptstyle\mathsf{H}}}

\newsavebox{\foobox}

\definecolor{kugray5}{RGB}{224,224,224}

\usepackage[normalem]{ulem}
\newcommand\rsout{\bgroup\markoverwith
	{\textcolor{red}{\rule[0.5ex]{2pt}{0.8pt}}}\ULon}



\makeatletter
\newcommand{\ALOOP}[1]{\ALC@it\algorithmicloop\ #1%
	\begin{ALC@loop}}
	\newcommand{\ENDALOOP}{\end{ALC@loop}\ALC@it\algorithmicendloop}

\makeatother

\usepackage{etoolbox}
\let\mybibitem\bibitem
\renewcommand{\bibitem}[1]{%
	\ifstrequal{#1}{nature}
	{\color{blue}\mybibitem{#1}}
	{\color{black}\mybibitem{#1}}%
}

\graphicspath{ {Figures/} }

\DeclareCaptionLabelSeparator{periodspace}{.\quad}

\addto\captionsenglish{}
\newcommand\nbthis{\addtocounter{equation}{1}\tag{\theequation}}

\newcommand{\norm}[1]{\left\lVert#1\right\rVert} % ||.||
\newcommand{\abs}[1]{\left|#1\right|} % ||
\newcommand{\tr}[1]{\mathrm{trace}\left(#1\right)} % ||
\newcommand{\re}[1]{\mathfrak{R}{\left(#1\right)}}

\allowdisplaybreaks
\newcommand{\mean}[1]{\mathbb{E} \left\{#1\right\}}

\usepackage{setspace}
\newcommand{\mR}{{\mathbf{R}}}
\newcommand{\mH}{{\mathbf{H}}}

\newcommand{\mW}{{\mathbf{W}}}
\newcommand{\mP}{{\mathbf{P}}}
\newcommand{\mI}{\textbf{\textbf{I}}}

\newcommand{\mC}{{\mathbf{C}}}

\newcommand{\mX}{{\mathbf{X}}}
\newcommand{\mY}{{\mathbf{Y}}}

\newcommand{\mF}{{\mathbf{F}}}
\newcommand{\mU}{{\mathbf{U}}}
\newcommand{\mV}{{\mathbf{V}}}

\newcommand{\setC}{\mathbb{C}} 

\newcommand{\vs}{{\mathbf{s}}}

\newcommand{\vy}{{\mathbf{y}}}

\newcommand{\vn}{{\mathbf{n}}}

\newcommand{\va}{{\mathbf{a}}}

\def\b0{{\pmb{0}}}

\newcommand{\Nr}{N_\text{r}}
\newcommand{\Nt}{N_\text{t}}

\newcommand{\Ns}{N_\text{s}}

\begin{document}

	\title{Joint Communications and Sensing Design for Multi-Carrier MIMO Systems}
	\author{\IEEEauthorblockN{Nhan~Thanh~Nguyen\IEEEauthorrefmark{1}, Nir Shlezinger\IEEEauthorrefmark{2}, Khac-Hoang Ngo\IEEEauthorrefmark{3}, Van-Dinh Nguyen\IEEEauthorrefmark{4}, Markku Juntti\IEEEauthorrefmark{1}}
		%% Affiliation
		\IEEEauthorblockA{\IEEEauthorrefmark{1}Centre for Wireless Communications, University of Oulu, P.O.Box 4500, FI-90014, Finland}
		\IEEEauthorblockA{\IEEEauthorrefmark{2}School of ECE, Ben-Gurion University of the Negev, Beer-Sheva, Israel}
		
		\IEEEauthorblockA{\IEEEauthorrefmark{3}Department of Electrical Engineering, Chalmers University of Technology, Gothenburg, Sweden}
		\IEEEauthorblockA{\IEEEauthorrefmark{4}College of Engineering and Computer Science, VinUniversity, Hanoi, Vietnam}\\
		Emails: \{nhan.nguyen, markku.juntti\}@oulu.fi; nirshl@bgu.ac.il; ngok@chalmers.se; dinh.nv2@vinuni.edu.vn}
	
	\maketitle

	\begin{abstract}
		In conventional joint communications and sensing (JCAS) designs for multi-carrier multiple-input multiple-output (MIMO) systems, the dual-functional waveforms are often optimized for the whole frequency band, resulting in limited communications--sensing performance tradeoff. To overcome the limitation, we propose employing a subset of subcarriers for JCAS, while the communications function is performed over all the subcarriers. This offers more degrees of freedom to enhance the communications performance under a given sensing accuracy. We first formulate the rate maximization under the sensing accuracy constraint to optimize the beamformers and JCAS subcarriers. The problem is solved via Riemannian manifold optimization and closed-form solutions. Numerical results for an $8 \times 4$ MIMO system with $64$ subcarriers show that compared to the conventional subcarrier sharing scheme, the proposed scheme employing $16$ JCAS subcarriers offers $60\%$ improvement in the achievable communications rate at the signal-to-noise ratio of $10$ dB. Meanwhile, this scheme generates the sensing beampattern with the same quality as the conventional JCAS design.
	\end{abstract}
	
	\begin{IEEEkeywords}
		Joint communications and sensing (JCAS), dual-functional radar-communications (DFRC), MIMO-OFDM.
	\end{IEEEkeywords}
	\IEEEpeerreviewmaketitle
	
	\section{Introduction}
	\label{sec:intro}
	The increasing demand for mobile communications services has raised the cost of bandwidth and caused frequency shortage \cite{johnston2022mimo}. Recently, cooperative spectrum sharing among licensed radar and communications systems has emerged as a key enabler for efficient use of bandwidth \cite{zhang2021overview}. Such coexisting systems are usually referred to as joint communications and sensing (JCAS) systems \cite{zhang2018multibeam}. Early studies on JCAS focused on interference management techniques to control the mutual interference between the radar and communications subsystems \cite{hassanien2016signaling, hassanien2015dual, liu2020joint, ma2020joint, chiriyath2017radar, zheng2019radar, li2016optimum, martone2019joint, liu2018mu, liu2017robust, buzzi2019using}. Particularly, Liu \textit{et al.} \cite{liu2018mu} proposed two JCAS operations for a multiple-input multiple-output (MIMO) downlink based on sharing array elements for the two subsystems. Hassanien \textit{et al.} \cite{hassanien2015dual} proposed transmitting communications data outside the main lobes of the radar. Wu \textit{et al.} \cite{wu2020waveform} introduced a frequency-hopping MIMO radar-based waveform for channel estimation. In \cite{liu2018toward, tang2022mimo, wu2023constant, liu2021dual}, constant-modulus waveforms were designed to avoid signal distortion in nonlinear power amplifiers.
	
	The aforementioned works mostly focus on single-carrier transmission. However, orthogonal frequency-division multiplexing (OFDM) can achieve a higher radar and communications performance \cite{keskin2021mimo, sturm2011waveform, ma2020joint}. In single-antenna OFDM systems \cite{ahmed2019ofdm, bicua2019multicarrier}, the major degrees of freedom is attained via power allocation \cite{ahmed2019ofdm} or dynamic subcarrier allocation \cite{bicua2019multicarrier}. Wu \textit{et al.} \cite{wu2022joint} optimized the MIMO-OFDM data symbols carried by subcarriers for better time and spatial domain signal orthogonality. Johnston \textit{el al.} \cite{johnston2022mimo} designed the radiated waveforms and the receive filters for MIMO-OFDM JCAS systems. More recent works \cite{temiz2021dual, cheng2021hybrid} focus on JCAS  designs in massive MIMO-OFDM systems. 
	
	Based on the subcarrier sharing among radar and communications, existing JCAS designs for multi-carrier/OFDM systems can be divided into two groups, wherein the two subsystems operate at either the same \cite{johnston2022mimo, wu2022joint, cheng2021hybrid, elbir2021terahertz, zhang2018multibeam} or distinct subcarriers \cite{bicua2019multicarrier}. Under sensing accuracy constraints, the former operation has communications performance loss across all the subcarriers, while that loss in the latter is even more significant due to the bandwidth fraction allocated for radar sensing. The higher accuracy required in sensing, the more significant performance loss occurs in communications. To overcome this, we herein propose an efficient JCAS design based on optimizing the subcarrier sharing and dual-functional beampatterns. The main idea is to allocate a subset of subcarriers for JCAS, while allowing communications over the entire bandwidth. In this way, radar sensing only interferes with data transmissions in predetermined sub-bands, while effective sensing can still be ensured with an efficient beampattern design. The communications via the remaining subcarriers are thus maximized without any effects from sensing. We formulate such a design as a rate maximization problem under sensing and power constraints. The problem is solved via subcarrier selection and Riemannian manifold optimization. Our numerical results show that the proposed JCAS scheme offers a remarkable improvement in the communications--sensing performance tradeoff with respect to the conventional counterpart.
	
	%\vspace{-0.15cm}
	\section{Signal Model and Problem Formulation}
	\label{sec_system_model}
	
	\subsection{Signal Model}
	We consider a MIMO-OFDM JCAS system, where the base station (BS) simultaneously transmits probing signals to the sensing targets at the angles of interest and data signals to the mobile stations (MS). Let $\Nt$ and $\Nr$ denote the numbers of antennas at the BS and MS, respectively, and let $\mathcal{K} = \{1,2,\ldots,K\}$ be the set of all employed subcarriers. Denote by $\vs[k] \in {\mathbb{C}}^{\Ns \times 1}$ the transmit signal vector at subcarrier~$k$, with $\mean{\vs[k] \vs[k]^\H}= \mI_{\Ns}$, where $\Ns$ is the number of data streams. We assume that while all subcarriers in $\mathcal{K}$ are used for communications, only a subset $\mathcal{J} \subset \mathcal{K}$ of $J = \abs{\mathcal{J}}$ subcarriers are used for sensing. Accordingly, subcarriers in $\mathcal{J}$ are employed for {\em both communications and sensing}, and thus, are referred to as \textit{JCAS subcarriers}.
	
	Let $\mF[k] \in \mathbb{C}^{\Nt \times \Ns}$ and $\mW[k] \in \setC^{\Nr \times \Ns}$ be the precoding matrix at the BS and the combining matrix at the MS, respectively, at subcarrier $k$. The power constraint at the BS is given as $\norm{\mF[k]}_{\mathcal{F}}^2 = P_{\text{BS}}, \forall k$, where $\norm{\cdot}_{\mathcal{F}}$ denotes the Frobenius norm, and $P_{\text{BS}}$ is the power budget at the BS. After combining, the received signal at the MS is given by 
	\begin{align*}
		\vy[k] &= \mW[k]^\H \mH[k] \mF[k] \vs[k] + \mW[k]^\H \vn[k], \nbthis \label{processed_received_signal}
	\end{align*}
	where $\vn[k] \sim \mathcal{CN}(\bm{0},\sigma^2_{\text{n}} \mI_{\Nr})$ is the additive white Gaussian noise~(AWGN) at the MS, and $\mH[k] \in \mathbb{C}^{\Nr \times \Nt}$ is the channel matrix at subcarrier~$k$. We assume that $\mH[k]$ is known at the BS and MS. Based on \eqref{processed_received_signal}, the achievable rate at subcarrier $k$ of the MS is given as
	\begin{equation}
		\!\!\mathcal{R}_k \!=\! \log_2 \abs{\mI_{\Nr}\! +\! \frac{P_{\text{BS}}}{\sigma^2_{\text{n}}  \Ns} \mW[k]^{\dagger} \mH[k] \mF[k] \mF[k]^\H \mH[k]^\H \mW[k]}, \! \label{eq_rate}
	\end{equation}
	where $\abs{\cdot}$ and $(\cdot)^{\dagger}$ denote the determinant and pseudo-inverse of a matrix, respectively.
	
	%\vspace{-0.15cm}
	\subsection{Problem Formulation} 
	
	\subsubsection{Radar Beampattern Design}
	The design of the radar beampattern is equivalent to the design of the covariance matrix of the radar probing signals, denoted as $\mR[k]$. This can be formulated as \cite{liu2018mu, liu2018toward}
	\begin{subequations}
		\label{opt_beampattern}
		\begin{align*} 
			&\hspace{-0.20cm}\underset{\substack{ \{ \mR[k] \} }}{\textrm{minimize}}\   \sum_{k \in \mathcal{J}} \sum_{t=1}^{T} \abs{\mathcal{P}_{\text{d}}(\theta_t,f_k)\hspace{-0.1cm} - \hspace{-0.1cm} \va(\theta_t,f_k)^\H  \mR[k]  \va(\theta_t,f_k)} \nbthis \label{obj_func_beampattern} \\
			&\hspace{-0.2cm} \textrm{subject to}\
			[\mR[k]]_{n,n} = P_{\text{BS}}/\Nt, \quad n \in \{1,\dots,\Nt\}, \nbthis \label{cons_avr_power} \\
			&\hspace{1.3cm}\mR[k] \succeq \bm{0}, \mR[k] = \mR[k]^\H,  \nbthis \label{obj_cons_beampattern}
		\end{align*}
	\end{subequations}
	where $\mathcal{P}_{\text{d}}(\theta_t,f_k)$ is the desired beampattern gain for angle $\theta_t$ and subcarrier $k$; $\{\theta_t\}_{t=1}^T$ defines a fine angular grid of $T$ angles covering the detection range $[-90^{\circ},90^{\circ}]$; $[\mR[k]]_{n,n}$ is the $n$-th diagonal element of $\mR[k]$ \cite{cheng2021hybrid, elbir2021terahertz}; and $\va(\theta_t, f_k) = [1, e^{j 2 \pi \frac{f_k \Delta}{c} \sin(\theta_t)}, \ldots, e^{j 2 \pi (\Nt-1) \frac{f_k \Delta}{c} \sin(\theta_t)}]^\T$
	is the steering vector of the BS, with $f_k$ and $\Delta$ being the is the $k$-th subcarrier frequency and the antenna spacing, respectively~\cite{cheng2021hybrid, johnston2022mimo}. Problem \eqref{opt_beampattern} is convex and can be solved by standard convex optimization tools such as CVX.
	
	\subsubsection{JCAS Design Problem}
	Let us denote $\mC[k] = \mF[k] \mean{\vs[k] \vs[k]^\H} \mF[k]^\H = \mF[k] \mF[k]^\H$ as the convariance matrix of the transmit signals at subcarrier $k$ in the JCAS system. The quality of the beampattern formed by $\mF[k]$ can be measured by the Frobenius norm $\norm{ \mR[k] - \mC[k] }_{\mathcal{F}}^2, \forall k \in \mathcal{J}$. Given $\mR[k]$ obtained via solving~\eqref{opt_beampattern} as the desired covariance matrix for radar sensing, we are interested in the JCAS beamforming design to $(i)$~maximize the system per-subcarrier achievable rate and $(ii)$~form beampatterns at subcarriers $k \in \mathcal{J}$ that match well with the radar beampattern $\va(\theta_t,f_k)^\H  \mR[k]  \va(\theta_t,f_k)$. This design is formulated as
	\begin{subequations}
		\label{opt_JCAS_design}
		\begin{align*} 
			\underset{\substack{ \{ \mF[k], \mW[k]\}_{k \in \mathcal{K}}, \mathcal{J} }}{\textrm{maximize}} \quad & \frac{1}{K} \sum_{k \in \mathcal{K}} \mathcal{R}_k \nbthis \label{obj_func} \\
			\textrm{subject to} \qquad
			& \norm{\mF[k]}_{\mathcal{F}}^2 = P_{\text{BS}},\ \forall k, \nbthis \label{cons_power}\\
			&\frac{1}{J} \sum_{k \in \mathcal{J}} \norm{\mR[k] - \mC[k] }_{\mathcal{F}}^2 \leq \tau_0, \nbthis \label{cons_beampattern}
		\end{align*}
	\end{subequations}
	where $\tau_0$ is a sensing accuracy tolerance. It is noted that the set $\mathcal{J}$ of integers is also a design variable.	The design in \eqref{opt_JCAS_design} is unlike most of the existing works on MIMO JCAS \cite{liu2018toward, liu2018mu}, which focus on maximizing sensing accuracy under the communications performance constraints. 
 
    Our communications-centric design \eqref{opt_JCAS_design} is motivated by the fact that when a large number of subcarriers is available, a subset of them can be used for sensing and guarantee a required accuracy without directly minimizing $\sum_{k \in \mathcal{J}} \norm{\mR[k] - \mC[k] }_{\mathcal{F}}^2$ as in \cite{liu2018toward, liu2018mu}. It offers more degrees of freedom to improve communications performance. Despite that, problem \eqref{opt_JCAS_design} is nonconvex and involves integer variables, i.e, $\mathcal{J}$. Our proposed solution to these challenges is elaborated next.
	
	\section{Proposed JCAS Design}
	
	We first note that eigenmode beamforming is not applicable for \eqref{opt_JCAS_design} due to constraint \eqref{cons_beampattern}. More specifically, setting $\mF[k]$ and $\mW[k]$ to the right and left singular vectors of $\mH[k]$, respectively, can maximize $\mathcal{R}_k$, but cannot ensure \eqref{cons_beampattern}. Instead, we first design the precoders $\{\mF[k]\}_{k \in \mathcal{K}}$ assuming that optimal combiners $\{\mW[k]\}_{k \in \mathcal{K}}$ are used.\footnote{In the following, we drop the subscript ${k \in \mathcal{K}}$ for ease of exposition.} Then, we optimize $\{\mW[k]\}$ for the designed $\{\mF[k]\}$. Furthermore, we exploit the observation that the sensing waveform constraint does not depend on the channels $\{\mH[k]\}_{k \notin \mathcal{J}}$, as seen from \eqref{cons_beampattern}. This implies that $\{\mF[k],\mW[k]\}$ can be designed to maximize the total rate $\sum_{k \in \mathcal{K} \backslash \mathcal{J}} \mathcal{R}_k$ over the set $\mathcal{K} \backslash \mathcal{J}$ without affecting the sensing performance. In this light, sensing has the smallest effect on communications if $\mathcal{J}$ contains subcarriers offering the smallest rates $\mathcal{R}_k$. This design strategy is outlined in the following steps:
	\begin{enumerate}
		\item Assuming optimal combiners $\{\mW[k]\}$, solve problem
		\begin{align} 
			\label{opt_JCAS_design_TX_comm}
			\underset{\substack{ \{ \mF[k] \} }}{\textrm{maximize}}\ \frac{1}{K} \sum_{k \in \mathcal{K}} \mathcal{R}_k,\ \textrm{subject to}\ \eqref{cons_power}
		\end{align}
		to obtain precoders $\{\hat{\mF}[k]\}$.
		\item Obtain set $\mathcal{J} \in \mathcal{K}$ containing $J$ subcarriers associated with the smallest rates $\mathcal{R}_k$.
		\item With given $\{\hat{\mF}[k]\}$, solve the JCAS problem
		\begin{subequations}
			\label{opt_JCAS_design_sensing}
			\begin{align*} 
				\underset{\substack{ \{ \mF[k]\}, k \in \mathcal{J} }}{\textrm{minimize}}\ &  {\frac{\rho}{J}} \sum_{k \in \mathcal{J}} \norm{ \mF[k] \mF[k]^\H - \mR[k]}_{\mathcal{F}}^2 \\
				& + {\frac{\bar{\rho}}{J}} \sum_{k \in \mathcal{J}}  \norm{\mF[k] - \hat{\mF}[k]}_{\mathcal{F}}^2 \nbthis \label{obj_func_JCAS} \\
				\textrm{subject to}\ & \eqref{cons_power} \nbthis \label{cons_power_TX_JCAS}
			\end{align*}
		\end{subequations}
		to obtain $\tilde{\mF}[k], k \in \mathcal{J}$, where $\rho$ is a weighting factor that balances the communications and sensing performance, and $\bar{\rho} = 1 - \rho$. Then, obtain the final precoders 
		\begin{align*}
			\mF[k] =
			\begin{cases*}
				\tilde{\mF}[k],\ k \in \mathcal{J}\\
				\hat{\mF}[k],\ \text{otherwise}.
			\end{cases*} \nbthis \label{eq_F_overall}
		\end{align*}
		\item With the obtained precoders $\mF[k]$, solve problem
		\begin{align*}
			\underset{\substack{ \{ \mW[k] \} }}{\textrm{maximize}} \quad  \frac{1}{K} \sum_{k \in \mathcal{K}} \mathcal{R}_k \nbthis \label{opt_JCAS_design_RX_comm}
		\end{align*}
		to obtain the optimal combiners $\{\mW[k]\}$.
	\end{enumerate}
	Note that in problem \eqref{opt_JCAS_design_sensing}, we have incorporated constraint \eqref{cons_beampattern} in the objective function as a penalty term. This is to simplify the design but does not reduce its efficiency because a better sensing accuracy can be always achieved by setting a larger $\rho$ in \eqref{obj_func_JCAS}, as will be further justified via simulations. Next, we present the solutions to subproblems \eqref{opt_JCAS_design_TX_comm}, \eqref{opt_JCAS_design_sensing}, and \eqref{opt_JCAS_design_RX_comm}.
	
	\subsubsection{Solution to Problem \eqref{opt_JCAS_design_TX_comm}}
	Let $\mV[k] \in \setC^{N_t \times N_s}$ be the matrix consisting of $N_s$ right singular vectors associated with the $N_s$ largest singular values of $\mH[k]$. Then, the optimal solution to $\mF[k]$ is given by
	\begin{align*}
		\mF[k] = \mV[k] \mP[k]^{1/2}, \nbthis \label{eq_solution_F}
	\end{align*}
	where $\mP[k]$ is a diagonal matrix with diagonal elements being water-filling power allocation factors satisfy \eqref{cons_power}.
	
	\subsubsection{Solution to Problem \eqref{opt_JCAS_design_sensing}}
	\label{sec_MO}
	We can rewrite \eqref{opt_JCAS_design_sensing} as
	\begin{align} 
		\label{opt_JCAS_design_sensing_1}
		\underset{\substack{ \{ \mF[k]\} }}{\textrm{minimize}} \quad {\frac{1}{J}} \sum_{k \in \mathcal{J}} \gamma_k,\ \textrm{subject to~}\ \eqref{cons_power},
	\end{align}
	where
	\begin{align*}
		\gamma_k &\triangleq \rho \norm{\mF[k] \mF[k]^\H - \mR[k]}_{\mathcal{F}}^2 + \bar{\rho} \norm{\mF[k] - \hat{\mF}[k]}_{\mathcal{F}}^2. \nbthis \label{eq_obj_MO}
	\end{align*}
	Since constraints \eqref{cons_power} for different $k \in \mathcal{J}$ are independent, problem \eqref{opt_JCAS_design_sensing_1} can be solved across $k$ independently, i.e.,
	\begin{align*}
		\underset{\substack{ \mF[k] \in \mathcal{S} }}{\textrm{minimize}} \quad \gamma_k, \nbthis \label{prob_MO}
	\end{align*}
	where $\mathcal{S} \triangleq \{\mF[k] \in \setC^{\Nt \times \Ns}: \norm{\mF[k]}_{\mathcal{F}} = \sqrt{P_{\text{BS}}}, \forall k \in \mathcal{J} \}$ is the complex hypershpere manifold with radius $\sqrt{P_{\text{BS}}}$. This motivates a Riemannian manifold minimization \cite{yu2016alternating} to efficiently find a near-optimal solution to \eqref{prob_MO}. To this end, we find the Euclidean gradient of $\gamma_k$ with respect to $\mF[k]$ as
	\begin{align*}
		\nabla \gamma_k \hspace{-0.1cm}=  4 \rho \left(\mF[k] \mF[k]^\H\hspace{-0.1cm} -\hspace{-0.1cm} \mR[k]\right) \mF[k] + 2 \bar{\rho} \big(\mF[k]\hspace{-0.1cm} -\hspace{-0.1cm} \hat{\mF}[k]\big). \nbthis \label{eq_gradient}
	\end{align*}
	A tangent space corresponds to $\mathcal{S}$ is given by
	\begin{align*}
		\mathscr{T}_{\mF[k]} \mathcal{S} = \left\{ \mF[k]  \in \setC^{\Nt \times M} : \re{ \tr{\mF[k]^\H \mF[k]} } = 0 \right\},
	\end{align*}
	where $\re{\cdot}$ denotes the real part of a complex number. We can obtain the Riemannian gradient corresponding to the Euclidean $\nabla \gamma_k$ by projecting $\nabla \gamma_k$ onto tangent space $\mathscr{T}_{\mF[k]} \mathcal{S}$, i.e.,
	\begin{align*}
		\mathrm{grad}\gamma_k &= \mathrm{Proj}_{\mF[k]} \left( \nabla \gamma_k \right)\\
		&= \nabla \gamma_k  - \re{ \tr{\mF[k]^\H \nabla \gamma_k} } \mF[k]. \nbthis \label{eq_projection}
	\end{align*}
	In an iteration of the Riemannian conjugate gradient scheme, say iteration  $i+1$, $\mF[k]$ is updated as follows:
	\begin{align*}
		\mF[k]_{(i+1)} &= \mathrm{Retract}_{\mF[k]_{(i)}} \big(\delta_{(i)} \bm{\Pi}_{(i)} \big) \\
		&= \frac{\sqrt{P_{\text{BS}}} \left(\mF[k]_{(i)} + \delta_{(i)} \bm{\Pi}_{(i)}\right)}{\norm{\mF[k]_{(i)} + \delta_{(i)} \bm{\Pi}_{(i)}}_{\mathcal{F}}} , \nbthis \label{eq_update_mQ}
	\end{align*}
	where $\delta_{(i)}$ is the Armijo step-size \cite{yu2016alternating, liu2018mu}, and $\bm{\Pi}_{(i)}$ is the descent direction. Here, $\bm{\Pi}_{(i)}$ is derived as
	\begin{align*}
		\bm{\Pi}_{(i)} = -\mathrm{grad}\gamma_{k{(i)}} + \mu_{(i)} \mathcal{T}_{\mF[k]_{(i-1)}}\big(\bm{\Pi}_{(i-1)}\big), \nbthis \label{eq_Pi_i}
	\end{align*} 
	where $\mathrm{grad}\gamma_{k{(i)}}$ is the Riemannian gradient $\mathrm{grad}\gamma_k$ at $\mF[k] = \mF[k]_{(i)}$, and $\mathcal{T}_{\mF[k]_{(i-1)}}(\bm{\Pi}_{(i-1)}) = \mathrm{Proj}_{\mF[k]_{(i)}}(\bm{\Pi}_{(i-1)})$ transports $\bm{\Pi}_{(i-1)}\in \mathscr{T}_{\mF[k]_{(i-1)}} \mathcal{S}$ to the tangent space $\mathscr{T}_{\mF[k]_{(i)}} \mathcal{S}$ using the projection defined in \eqref{eq_projection}, and $\mu_{(i)}$ is computed as 
	\begin{align*}
		\mu_{(i)} = \frac{\langle \mathrm{grad}\gamma_{k{(i)}}, \overline{\mathrm{grad}}\gamma_{k{(i)}} \rangle}{\langle \mathrm{grad}\gamma_{k{(i-1)}}, \mathrm{grad}\gamma_{k{(i-1)}} \rangle}, \nbthis \label{eq_mu_i}
	\end{align*}
	with $\langle \mX, \mY \rangle \triangleq \re{\tr{\mX^\H \mY}}$ and $\overline{\mathrm{grad}}\gamma_{k{(i)}}$ $\triangleq\mathrm{grad}\gamma_{k{(i)}} - \mathscr{T}_{\mF[k]_{(i-1)}}\left(\mathrm{grad}\gamma_{k{(i)}}\right).$
	
	\subsubsection{Solution to Problem \eqref{opt_JCAS_design_RX_comm}}
	Given $\mF[k]$, the optimal solution to $\mW[k]$ is given as
	\begin{align*}
		\mW[k] = \mU[k], \nbthis \label{eq_sol_W}
	\end{align*}
	where $\mU[k]$ consists of $\Ns$ left singular vectors corresponding to the $\Ns$ largest singular values of $\mH[k] \mF[k]$.
	
	\subsubsection{Overall Proposed JCAS Design}
	The proposed design for the considered JCAS problem is outlined in Algorithm~\ref{alg_MO}. Specifically, $\{\hat{\mF}[k]\}$ are first obtained in step 1 for the maximum communications rates via all subcarriers. In step 2, we solve the benchmark covariance matrices $\{\mR[k]\}$, which are then used in steps 4--14 to find  $\{\Tilde{\mF}[k]\}$ based on the Riemannian manifold optimization detailed in Section \ref{sec_MO}. The final beamformers and combiners are obtained in steps 15 and 16, respectively.%$\{\mW[k]\}$ are obtained in step 14.
 
    \begin{algorithm}[t]
		\small
		\caption{Proposed solution to the JCAS design~\eqref{opt_JCAS_design}}
		\label{alg_MO}
		\begin{algorithmic}[1]
			\REQUIRE $\{\mH[k], \mathcal{P}_{\text{d}}(\theta_t,f_k), \va(\theta_t,f_k)\}$.
			\ENSURE $\left\{ \mF[k], \mW[k] \right\}$.
			\STATE Obtain $\{\mF[k]\}$ based on \eqref{eq_solution_F}. Set $\hat{\mF}[k] = \mF[k], \forall k \in \mathcal{K}$.
			\STATE Find the set $\mathcal{J}$ of $J$ subcarriers offering the smallest rates.
			\STATE Solve problem \eqref{opt_beampattern} to obtain $\{\mR[k]\}$.
			\FOR{$k \in \mathcal{J}$}
			%\STATE Obtain $\hat{\mF}$ based on \eqref{def_Qbar} with $k = \mathcal{J}\{j\}$.
			\STATE Set $i=0$, initialize $\mF[k]_{(0)}$ and $\bm{\Pi}_{(0)} = -\mathrm{grad} \gamma_{k{(0)}}$.
			\REPEAT
			\STATE Compute stepsize $\delta_{(i)}$ by Armijo rule.
			\STATE Update $\mF[k]_{(i)}$ based on \eqref{eq_update_mQ}.
			\STATE Compute $\mu_{(i)}$ based on \eqref{eq_mu_i}.
			\STATE Compute $\bm{\Pi}_{(i)}$ based on \eqref{eq_Pi_i}.
			\STATE $i = i + 1$.
			\UNTIL{convergence}
			\ENDFOR
			\STATE Set $\tilde{\mF}[k]$ to the solutions $\mF[k]_{(i)}$ at convergence, $k \in \mathcal{J}$.
			\STATE Obtain final beamformers $\{\mF[k]\}$ based on \eqref{eq_F_overall}.
			\STATE Obtain combiners $\{\mW[k]\}$ based on \eqref{eq_sol_W}.
		\end{algorithmic}
	\end{algorithm} 
	
    \subsubsection{Complexity Analysis} 
	We end this section with a complexity analysis of the proposed algorithm. The complexity of obtaining $\{\mF[k]\}$ in \eqref{eq_solution_F} is $K \mathcal{O}(\Nt\Nr^2)$, dominated by the singular value decomposition (SVD) of $\{\mH[k]\}$. The complexity of the Riemannian manifold minimization is $J \mathcal{O}(4\Nt^2 \Ns + 3\Nt)$, which is mostly to compute the Euclidean gradient in~\eqref{eq_gradient}~\cite{liu2018mu}. Finally, solving $\{\mW[k]\}$ requires performing multiplication and SVD of $\{\mH[k] \mF[k]\}$ with a complexity of $K \mathcal{O}(2\Nt \Nr \Ns + \Nr^2 \Ns)$. Therefore, the overall complexity of Algorithm \ref{alg_MO} is $\mathcal{O}(4J\Nt^2\Ns + K \Nt \Nr^2 + 2K  \Nt\Nr\Ns + K \Nr^2\Ns)$.

	\section{Simulation Results}
	
	\begin{figure}[t]
		\includegraphics[scale=0.599]{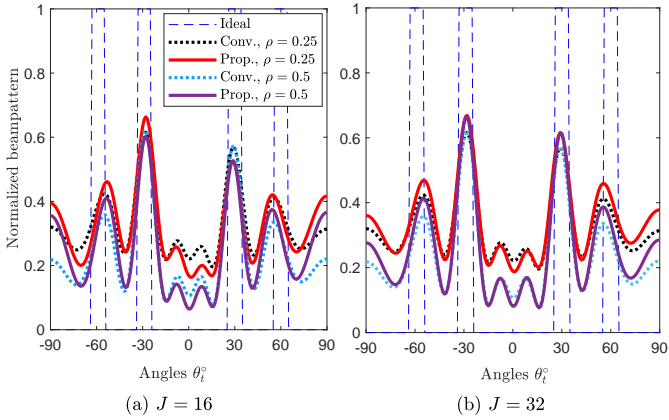}
		\caption{Beampatterns of the proposed and conventional JCAS schemes with $\rho = \{0.25,0.5\}$, $J = \{16,32\}$, and SNR $=10$ dB.}% It should be made explicite that Prop. refers to our scheme and Conv. refers to $J=K$.}
	\label{fig_beam_vs_angle}
\end{figure}

\begin{figure}[t]
	\includegraphics[scale=0.555]{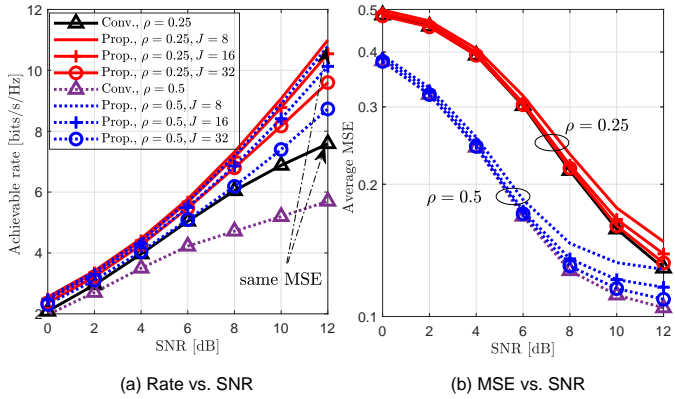}
	\caption{Achievable rates and MSEs of the proposed and conventional JCAS schemes with $\rho = \{0.25,0.5\}$ and $J = \{8,16,32\}$.}
	\label{fig_rate_MSE_vs_SNR}
\end{figure}

We herein provide numerical results to demonstrate the efficiency of the proposed JCAS design. In the simulations, we set $\Nt = 8$, $\Nr = 4$, $K = 64$, $\sigma_{\text{n}}^2 = 1$, and $f_k = f_0 + (k-1) \Delta f$, where $f_0 = 2$ GHz and $\Delta f = 100$ kHz \cite{johnston2022mimo}. We assume the deployment of a uniform linear array (ULA) with antenna spacing $\Delta = c/(2 f_K)$~\cite{johnston2022mimo}, where $c \approx 3\times 10^8$~m/s is the speed of light. We adopt the Rayleigh fading model for the channels~$\mH[k], \forall k \in \mathcal{K}$~\cite{liu2018mu}. The sensing targets are assumed to be located at angles $\theta_{\text{d}} \in \{-60^{\circ}, -30^{\circ}, -30^{\circ} , 60^{\circ}\}$, and the corresponding desired beampattern is defined as \cite{cheng2021hybrid}
\begin{align*}
	\mathcal{P}_{\text{d}}(\theta_t,f_k) = \begin{cases}
		1,\ \theta_t \in [\theta_{\text{d}} - \delta_{\theta}, \theta_{\text{d}} + \delta_{\theta}]\\
		0,\ \text{otherwise}
	\end{cases},\ \forall k, \nbthis \label{eq_ideal_beampattern}
\end{align*}
where $\delta_{\theta} = 8$ is the half of the mainlobes of $\mathcal{P}_{\text{d}}(\theta_t,f_k)$. The Riemannian manifold optimization is implemented using the Manopt toolbox \cite{manopt}. All the results are averaged over $100$ channel realizations. In the figures, the term ``Prop.'' refers to the proposed JCAS design with $J < K$, while the conventional design with $J = K$ is referred to as ``Conv.''

In Fig.\ \ref{fig_beam_vs_angle}, we plot the beampatterns obtained by the proposed JCAS design with $J = \{16, 32\}$ in comparison to those of the conventional setup, i.e., $J=K$, as well as the ideal beampattern in \eqref{eq_ideal_beampattern}. For problem \eqref{opt_JCAS_design_sensing}, we consider moderate weighting factors of $\rho = \{0.25, 0.5\}$. % for sensing part, and the correponding weights for communications are $\Bar{\rho} = \{0.75, 0.5\}$. 
It is observed that despite the reduced number of JCAS subcarriers, the proposed scheme still forms good beampatterns with peaks located at desired detection angles $\theta_{\text{d}}$, and they are as high as those of the conventional beampatterns. Around the margin angles $\{\pm 90^{\circ}\}$, the proposed beampatterns exhibit a higher error than the conventional ones, which can be reduced by increasing $J$, as seen for the case $J = 32$. When $\rho$ increases, the power distributes less over sidelobe angles $\theta_t \notin \{-60^{\circ}, -30^{\circ}, -30^{\circ} , 60^{\circ}\}$. This enhances the peak-to-sidelobe ratio (PSLR) and improves the sensing accuracy \cite{liu2018mu, liu2018toward} but essentially degrades the communications performance, as shown next.

In Fig.~\ref{fig_rate_MSE_vs_SNR}, we compare the achievable communications rates and the average sensing beampattern mean squared errors (MSEs) of the considered JCAS schemes versus SNRs (defined as $P_{\text{BS}}/{\sigma_{\text{n}}^2}$) for $\rho = \{0.25, 0.5\}$ and $J = \{8,16, 32\}$. Here, the average beampattern MSE is defined as $\text{MSE} = \frac{1}{JT}\sum_{k \in \mathcal{J}} \sum_{t=1}^{T} \abs{\mathcal{P}_{\text{d}}(\theta_t,f_k) - \va(\theta_t,f_k)^\H  \mC[k]  \va(\theta_t,f_k)}^2$. It is seen that for the same $\rho$, the proposed JCAS design achieves remarkable rate improvement with only a marginal loss in the MSE. In particular, the improvement is more significant with a larger $\rho$, which is reasonable because the proposed subcarrier sharing plays a more important role under strict sensing constraints. This implies that employing a larger $\rho$ and a smaller $J$ should be employed for better communication--sensing performance tradeoff. As an example, at SNR~$=12$ dB, the proposed scheme with $\rho = 0.5$ and $J = 8$ achieves $42.4\%$ higher rate while maintaining the same MSE ($=0.128$) as the conventional scheme with $\rho=0.25$ and $J=K$.

\begin{figure}[t]
    \hspace{-0.25cm}
	\includegraphics[scale=0.65]{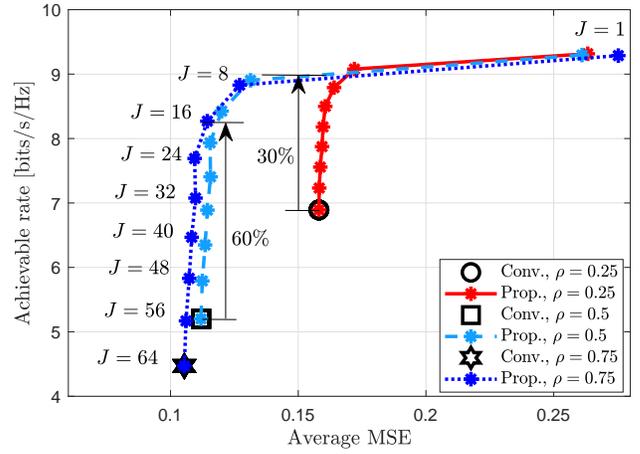}
	\caption{Communications-sensing performance tradeoff of the proposed JCAS scheme with $\rho = \{0.25,0.5,0.75\}$, $J = \{1,8,16,\ldots,K\}$, SNR $=10$ dB.}
	\label{fig_tradeoff}
\end{figure}

We further demonstrate the superior tradeoff improvement of the proposed scheme in Fig.\ \ref{fig_tradeoff} by showing the communications rate versus the MSEs obtained for $J = \{1,8,16,\ldots,64\}$, $\rho = \{0.25, 0.5, 0.75\}$, and SNR $=10$ dB. It is seen for the proposed scheme that as $J$ increases, both the MSEs and communications rates decrease to reach those of the conventional one. However, for a sufficiently large $J$, e.g., $J \geq 16$, the curves of the proposed scheme almost align with vertical lines along the y-axis. This clearly shows that the proposed JCAS design remarkably improves the communications rates while maintaining the sensing performance. For example, with $J=16$, $\rho = 0.75$, and for the same sensing MSE~$=0.11$, the proposed scheme offers an improvement of $60\%$ in rate performance as the conventional scheme with $J= K$ and $\rho = 0.0.5$.

\section{Conclusion}

We have investigated JCAS design in a MIMO-OFDM system. Aiming at maximizing the achievable communications rate while ensuring sensing accuracy, we have proposed an efficient beamforming design with a new subcarrier allocation strategy. Specifically, we proposed to use a subset of subcarriers for radar sensing, while leveraging all the available subcarriers for communications. Numerical results show that reliable sensing beampatterns can still be achieved with a reduced number of subcarriers, while the communications performance is dramatically improved. Equivalently, the communications-sensing performance tradeoff has been improved remarkably.

\section*{Aknowledgement}
This research was supported by Academy of Finland under 6G Flagship (grant 346208), Infotech Oulu, and Business Finland, Keysight, MediaTek, Siemens, Ekahau and Verkotan via project 6GLearn.

% \newpage
\bibliographystyle{IEEEtran}
\balance
\bibliography{IEEEabrv,Bibliography}
\end{document}